\documentclass[doublecol,a4paper,showpacs,superscriptaddress,floatfix]{epl2} 

\usepackage{graphicx}
\usepackage{amsmath}
\usepackage{bm}
\usepackage{amssymb}
\usepackage[vcentermath]{youngtab}

\newcommand{\SU}{SU$_4$ }
\newcommand{\bpi}{\bm{\pi}}

\begin{document}

\title{Symmetry content and spectral properties of charged collective excitations
 for graphene in strong magnetic fields
       }
\shorttitle{Charged collective excitations
 for graphene in strong magnetic fields}
\author{A.M.\ Fischer\inst{1} \and R.A.\ R\"omer\inst{1} \and A.B.\ Dzyubenko\inst{2,3}}
\shortauthor{A.M.\ Fischer \etal}
\institute{
\inst{1} Department of Physics and Centre for Scientific Computing,
  University of Warwick, Coventry CV4 7AL, United Kingdom\\
\inst{2} Department of Physics, California State University Bakersfield, Bakersfield, CA 93311, USA\\
\inst{3} General Physics Institute, Russian Academy of Sciences, Moscow 119991, Russia
}

\date{$Revision: 1.16 $, compiled \today}

\abstract{
We show that graphene in a strong magnetic field with partially filled Landau levels sustains charged collective excitations --- bound states of a neutral magnetoplasmon and free particles. In the limit of low density of excess charges, these are bound three-particle complexes.
Some of these states are optically bright and may be detected in spectroscopy experiments,
providing a direct probe of electron-electron interactions in graphene.
The charged excitations can be classified using the geometrical symmetries --- non-commutative magnetic translations
and generalized rotations --- in addition to the dynamical \SU symmetry in graphene. From the \SU symmetry point of view,
such excitations are analogous to bound states of two quarks and one antiquark $qq\bar{q}$
with four flavors. We establish a flavor optical selection rule to identify the bright states for experimental studies.
}


\maketitle
\mbox{}
 \vspace{-20pt}
 \mbox{}

Since the isolation of graphene, the unusual behavior of its electrons
in a magnetic field has been the subject of intense study \cite{CasGPNG09,AbeABZ10}.
Early on, the anomalous integer quantum Hall effect was observed,
with plateaus detected at filling factors $\nu=\pm2\left(|n|+1\right)$,
where $n$ is the Landau level (LL) index \cite{NovGMJ05,ZhaTSK05}.
Subsequently, in higher fields,
plateaus were seen at $\nu = 0,\pm1, \pm4$, signifying a lift in the fourfold (spin and valley)
LL degeneracy, most likely due to the many-body effects \cite{JiaZSK07}.
Recently, fabrication of high quality samples has allowed
the fractional quantum Hall effect (FQHE) to be observed at $\nu=\frac{1}{3}$,
indicating again
the presence of a strongly correlated electron state \cite{DuSDL09,BolGSS09}.
Importantly, such signatures of {\em correlated many-body physics} emerge already at low magnetic fields, and in higher fields these correlations determine the physics.
Here we establish the existence of {\em collective charged excitations}
which are essentially bound states of a neutral magnetoplasmon binding free particles.
In the limit of low density of excess carriers such collective excitations
can be thought of as three-particle interacting states.
We will demonstrate that in graphene complexes of electrons,
akin to $qq\bar{q}$ quark-triplets in high-energy physics, arise naturally.
This is because the spin and valley pseudospin in graphene combine
to form four {\em flavors} analogous to the four flavors of 1st and 2nd generation quarks \cite{GreM97,quarks}.
Analysis of the symmetry content of these states allows the classification of
optically active and dormant states, highlighting the usefulness of high energy
analogues in graphene.

\begin{figure}[t]
  \centering
  \includegraphics[width=0.82\columnwidth]{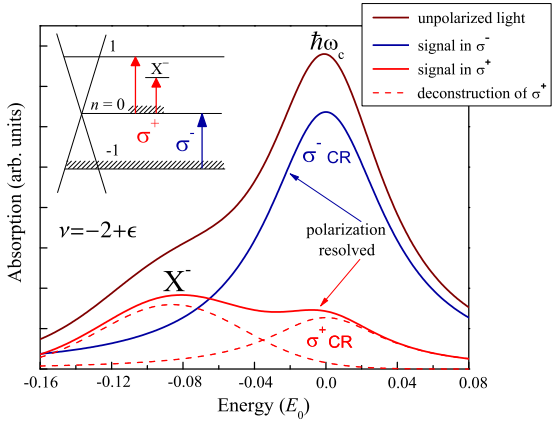}
  \caption{(color online) Predicted absorption spectra in pristine graphene with \mbox{$\nu = -2 + \epsilon$}
  for the bright state at energy \mbox{$-0.09 E_0$}.
  The inset schematically shows the relevant optical transitions.
  Energies are given in units of $E_0$ (see text) with zero energy corresponding to the CR energy, $\hbar\omega_c$.
    }
  \label{Optics}
\end{figure}

We study the formation of such excitations in pristine graphene
in strong magnetic fields with a low density of excess electrons in partially filled LL's.
More specifically, we consider inter-LL excitations which, in the presence of excess charges,
may bind one additional electron (or hole).
Such excitations (i) lie above charged topological collective excitations,
Skyrmions \cite{AroKL99,EzaT05,YanDM06}, and (ii) are directly accessible via optical and tunneling experiments.
In particular, we are interested in the behavior of graphene close to the charge neutrality point
and thus focus on filling factors $\nu = \mu \pm \epsilon$ for $\mu=0, \pm1,\pm2$ and $\epsilon \ll 1$.
By way of illustration, let us consider the filling factor $\nu=-2+\epsilon$,
so that the $n=0$ LL is almost empty (see Fig.~\ref{Optics}, inset).
A neutral collective excitation, or magnetoplasmon, is formed when an electron,
excited to an empty state in the $n=0$ LL, becomes bound to the hole formed in the $n=-1$ LL.
On their own, such neutral excitations form a continuum of {\em extended} states,
with the lower continuum edge corresponding to the cyclotron resonance
(CR) mode at energy $\hbar\omega_c =  v_F \sqrt{\frac{2 e \hbar B}{c}}$ \cite{JiaHTW07}.
In addition, they may bind one of the few excess electrons in the $n=0$ LL, thus forming a negatively charged
trion $(eeh)$ complex, $X^-$.
These {\em discrete} bound states are stable when located below the magnetoplasmon continuum.
We demonstrate the existence of such states, and determine their spectral properties.
We find that the symmetry of graphene severely restricts optical transitions
so that most collective states are dark.
The bright bound states give rise to additional optical absorption
peaks below the CR as shown in Fig.~\ref{Optics} and whose observation is within experimental reach.
The binding energies of the extra electron in the $X^-$ are $\sim 0.1 E_0$,
similar to large FQHE gaps expected in pristine graphene \cite{ApaC06,AbaSDA10} ($E_0 \simeq 40$\,meV at $B=10$\,T).
Here
$E_\mathrm{0}= \sqrt{\frac{\pi}{2}} \frac{e^2}{\varepsilon l_B}$
is the characteristic energy of Coulomb interactions
in strong fields, $E_\mathrm{0} \ll \hbar\omega_c $;
$l_B =\left( \hbar c/eB\right)^{1/2}$ is the magnetic length and $\varepsilon$ the effective dielectric constant.
In this regime, virtual transitions between LL's are suppressed as powers of
$E_\mathrm{0}/\hbar\omega_c$ and are neglected, constituting the high magnetic field approximation
 \cite{YanDM06,GoeMD06,ApaC06,IyeWFB08,FisDR09a}.

Due to the particle-hole symmetry in graphene,
positively charged trions $X^+$, or $(ehh)$ states,
exist as $X^-$ counterparts. This occurs for the complementary filling factors $-\nu$,
such as, $-\nu=2-\epsilon$, when the $n=0$ LL is almost completely filled but contains a few holes.
The $X^+$ states at $-\nu$ have the same energies and optical strengths as their $X^-$ counterparts
at $\nu$, but are active in the opposite circular polarization.

\begin{figure*}[t]
  \centering
   \includegraphics[width=0.69\columnwidth]{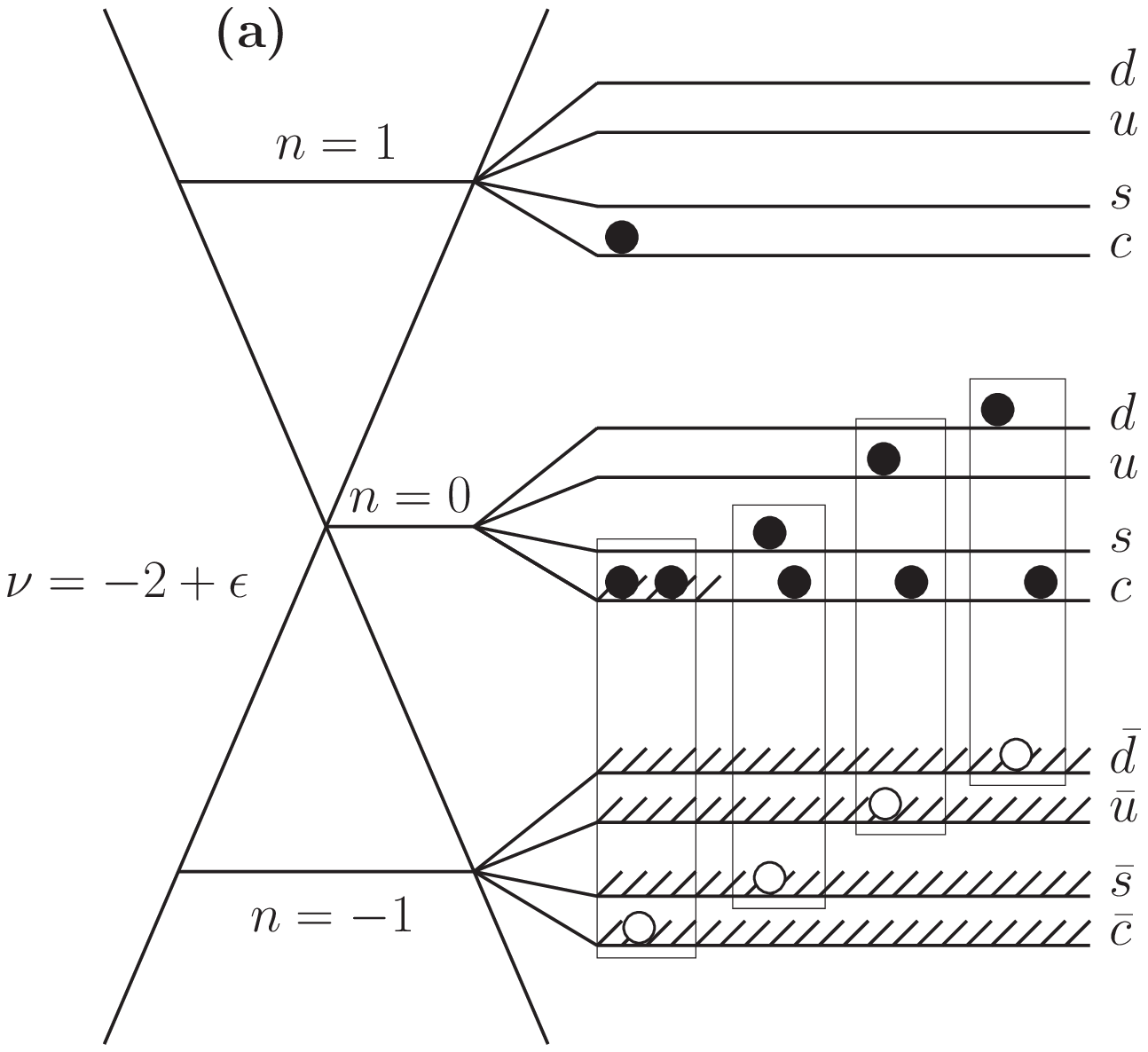}
  \includegraphics[width=0.65\textwidth]{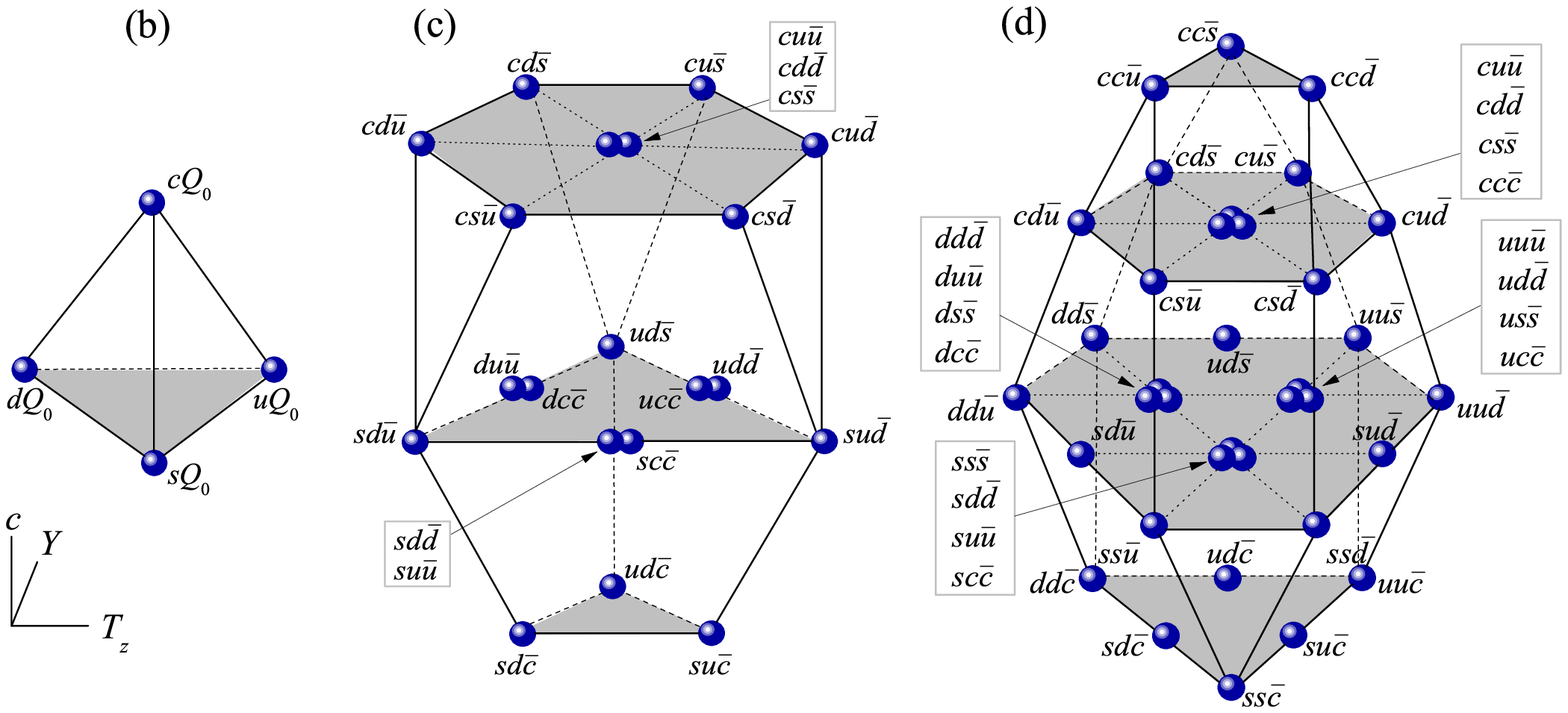}
  \caption { (color online)
  (a) Filled/empty circles ($\bullet$/$\circ$) denote electrons/holes; hatched regions, the Dirac sea.
	  States in the boxes are resonantly mixed by the Coulomb interactions.
    (b) Quartet $\left[4 \right]$ and  (c) $\left[ \smash{\overline{20}} \right]$,
  and (d) $\left[ 36\right]$ SU$_4$ multiplets
  describe the trion $X^-$ states. Shading denotes the different levels of charm.
  The neutral $(eh)$ state in (b)
  $Q_0 = \frac{1}{2} \left( d\bar{d} + u\bar{u} +  s\bar{s} + c\bar{c} \right)$ is the \SU singlet.
  The electron in the $n=1$ LL shown in (a) is resonantly admixed to the $X^-$ from the $[4]$-multiplet with $J_z = \frac{1}{2}$.
}
\label{fig-bound}
\end{figure*}

With the underlying \emph{dynamical} \SU symmetry of graphene,
resulting from two possible spin ($\uparrow,\downarrow$)
and two valley pseudospin ($\Uparrow,\Downarrow$) projections,
the trion states are in fact completely analogous, from the symmetry point of view,
to bound states of two quarks and one antiquark with four flavors, $qq\bar{q}$ for the $X^-$
and $q\bar{q}\bar{q}$ for the $X^+$ \cite{quarks}.
The corresponding multiplets are shown in Fig.~\ref{fig-bound};
LL splittings in Fig.~\ref{fig-bound}(a) are shown for clarity.
In particular, the state designated in Fig.~\ref{Optics} as $X^-$ realizes the fundamental
representation of the \SU group, Fig.~\ref{fig-bound}(b).
The elementary particle physics classification of states and
the techniques of the \SU group \cite{GreM97} turn out to be very useful in graphene.
They allow us to perform a complete group-theoretical analysis of the bound states and
to determine the multiplicities of the emerging groups of excitations. Importantly, these enable us to establish a flavor optical selection rule, explaining which of the states are dark and which are bright for experimental studies.

In the following, we focus on the $X^-$ states, which may occur for fillings $\nu = \mu + \epsilon$,
where $\epsilon \ll 1$ and $\mu$ is an integer; the $X^+$ case proceeds analogously.
The operator creating a charged collective excitation is
\begin{equation}
  \label{eq-Rdagger} 
 \mbox{} \hspace{-3pt}
      R^{\dag}_{k J_z \mathbf{N}}= \!\!\!\!
          \sum_{m_1, m_2 = 0}^{\infty}
           A_{k J_z \mathbf{N}}(m_1,m_2) \,
            c^{\dag}_{N_1 m_1}
			c^{\dag}_{N_2 m_2}
			d^{\dag}_{N_3 m_3} \; ,
\end{equation}
where the hole representation $d^{\dag}_{N m} \equiv c_{N m}$
is used for all completely filled levels.
The composite index $N = \{ n, s_z, \sigma_z \} $ carries the LL index $n$,
the spin $s_z=\pm\frac{1}{2}$ and the pseudospin  $\sigma_z=\pm\frac{1}{2}$
(labeling valleys $K$ and $K'$), while $\mathbf{N}=\{ N_1,N_2,N_3 \}$.
The oscillator quantum number, $m$, represents in the symmetric gauge $\mathbf{A}= \frac{1}{2} \mathbf{B} \times \mathbf{r}$
the guiding center for a single particle, yielding the LL degeneracy.
As for the neutral case \cite{FisDR09a}, several of the $R^{\dag}_{k J_z \mathbf{N}}$ states having same $k$, $J_z$
and equal or close energies (but different $\mathbf{N}$'s) are strongly mixed by Coulomb interactions.
As a result, the true excitation is given by a linear combination of these (see Fig.~\ref{fig-bound}).

The {\em geometrical} symmetries of the system allow the classification of the composite excitations
by two orbital quantum numbers,
half integer $J_z$ and integer  $k=0, 1, \ldots$.
The former is due to the symmetry under generalized rotations in graphene. It corresponds
to the total $\hat{J}_z = \sum_i \hat{j}_{iz} $, where a single-particle operator
$\hat{j}_{iz} = \hat{l}_{iz} - \frac{1}{2} {\bf 1}_2 \otimes \tau_{iz}$
involves the orbital $\hat{l}_{iz}$
and the sublattice isospin $\tau_{iz}$ angular momentum  projections \cite{DiVM84,FisDR09a}.
The oscillator quantum number, $k$, is due to magnetic translations whose generator
$\hat{\mathbf{K}} = \sum_i \hat{\mathbf{K}}_i$ commutes with the Hamiltonian $\hat{H}$.
The single particle operators are
$\hat{\mathbf{K}}_i = \hat{\bpi}_i - \frac{e_i}{c}\hat{\mathbf{r}}_i \times \mathbf{B}$,
where $\hat{\bpi}_i$
is the kinematic momentum operator and $e_i=\pm e$ the particle's charge.
For charged complexes, the components do not commute: $[\hat{K}_x,\hat{K}_y]= i \hbar eB/c$ for the $X^-$.
However, using the mutual commutativity of $\hat{H}$, $\hat{J}_z$, and $\hat{\mathbf{K}}^2$,
the composite states
can be chosen to be eigenstates of $\hat{\mathbf{K}}^2$
with eigenvalues $(2k+1)(\hbar/l_B)^2$ and $k = 0, 1, \ldots$ \cite{DzyS00}.
Physically, the oscillator number $k$ determines the distance
at which the guiding center of a charged complex as a whole is located relative to the origin.
For a uniform system this leads to the Landau degeneracy in $k$.
Therefore, charged composites occur in families starting with a \emph{seed} state, which has
$k=0$ and a certain value of $J_z$; the latter cannot be guessed {\it a priori}.
The offspring $k = 1, 2, \ldots$ are generated by successive action of
the raising ladder operator $\hat{K}_{-}= \frac{l_B}{\sqrt{2}\,\hbar}\left( \hat{K}_x - i \hat{K}_y\right)$. 
They have decreasing eigenvalues of $\hat{J}_z$ equal to
$J_z - k$ because of the relation $\left[ \hat{J}_z,\hat{K}_{-} \right] = - \hat{K}_{-}$.
In addition to having the same energies as each other,
states in a given family also exhibit the same optical properties \cite{DzyNYM01}.

The Hamiltonian is given by $\hat{H}=\hat{H}_{0} + \hat{H}_{\rm int} $
with the interaction part $\hat{H}_{\rm int} = \hat{H}_{ee} + \hat{H}_{hh} + \hat{H}_{eh}$
and
$
 \hat{H}_{0} = 
     \sum_{ \substack{N , m \\ \tilde{\epsilon}_N >  \epsilon_{\rm F} } }
	               \tilde{\epsilon}_N c^{\dag}_{N m} c^{\vphantom{\dag}}_{N m}
   - \sum_{ \substack{N , m \\ \tilde{\epsilon}_N \leq  \epsilon_{\rm F} } }
                   \tilde{\epsilon}_N d^{\dag}_{N m} d^{\vphantom{\dag}}_{N m}
$
being the noninteracting part.
Note that the range over which a given $N$ quantum number is summed,
depends on whether it indexes an electron $c^{\dag}_{N m}$ or hole $d^{\dag}_{N m}$ operator.
The single-particle energy $\tilde{\epsilon}_{N} = \epsilon_{N} + E_{\rm SE}(N)$
denotes the bare LL energy $\epsilon_{N}$
renormalized by $e$-$e$ exchange self-energy corrections $E_{\rm SE}(N)$  \cite{IyeWFB08,FisDR09a,FisRD10}.
The bare LL energies $\epsilon_{N} = {\rm sign}(n) \hbar\omega_c \sqrt{|n|} + \hbar\omega_s s_z + \hbar\omega_{v} \sigma_z$
may phenomenologically include the Zeeman splitting $\hbar\omega_s$ and a possible valley splitting $\hbar\omega_{v}$;
these are assumed to be small, $\hbar\omega_{s,v} \ll E_0$.
The electron-electron  $\hat{H}_{ee}$, hole-hole $\hat{H}_{hh}$, and electron-hole $\hat{H}_{eh}$
parts of the interaction Hamiltonian $\hat{H}_{\rm int}$ are found by projecting onto particular LL's.
Their forms for excitations at $\nu = -2 + \epsilon$
are (cf.\ Fig.~\ref{fig-bound}(a))
\begin{equation}
\label{Hee} 
 \hat{H}_{ee} = \frac{1}{2}
        \sum_{\substack{m_1 , m_2 \\ m_1' , m_2'}}
	    \sum_{\substack{s_1 , \sigma_1 \\ s_2 , \sigma_2}}
         \mathcal{W}_{N_1 m_1  \, N_2 m_2}^{N_1' m_1' \, N_2' m_2'}
                c^{\dag}_{N_1' m_1'} c^{\dag}_{N_2' m_2'} c^{\vphantom{\dag}}_{N_2 m_2} c^{\vphantom{\dag}}_{N_1 m_1} 				
\end{equation}
with $N_i=N_i'=\{0, s_i, \sigma_i \}$ and $i=1,2$.
The graphene Coulomb interaction vertex
$\mathcal{W}_{N_1 m_1  \, N_2 m_2}^{N_1' m_1' \, N_2' m_2'} \sim
\delta_{s_1,s_1'} \delta_{\sigma_1,\sigma_1'}
\delta_{s_2,s_2'} \delta_{\sigma_2,\sigma_2'}$ conserves spin and pseudospin and
may be expressed in terms of the vertices calculated using the standard 2D electron wavefunctions in LL's \cite{FisDR09a}.
The Hamiltonian of the \mbox{$e$-$h$} interactions consists of two parts,
$\hat{H}_{eh} =  \hat{H}_{eh}^{\rm I} + \hat{H}_{eh}^{\rm II}$, where
\begin{equation}
\label{HeheI} 
 \hat{H}_{eh}^{\rm I} =
        \sum_{\substack{m_1 , m_2 \\ m_1' , m_2'}}
	    \sum_{\substack{s_1 , \sigma_1 \\ s_2 , \sigma_2}}
         \bar{\mathcal{W}}_{N_1 m_1  \, N_2 m_2}^{N_1' m_1' \, N_2' m_2'}
                c^{\dag}_{N_1' m_1'} d^{\dag}_{N_2' m_2'} d^{\vphantom{\dag}}_{N_2 m_2} c^{\vphantom{\dag}}_{N_1 m_1} 
\end{equation}
with $N_1=N_1'=\{0, s_1, \sigma_1 \}$ and   $N_2=N_2'=\{-1, s_2, \sigma_2 \}$.
The \mbox{$e$-$h$} vertex
%
%
\begin{equation}
   \label{Weh} 
        \bar{\mathcal{W}}_{N_1 m_1  \, N_2 m_2}^{N_1' m_1' \, N_2' m_2'}  =
             \mathcal{W}_{N_1 m_1  \, N_2' m_2'}^{N_1' m_1' \, N_2 m_2} -
             \mathcal{W}_{N_1 m_1  \, N_2' m_2'}^{N_2 m_2 \, N_1' m_1'}
\end{equation}
describes the direct \mbox{$e$-$h$} Coulomb attraction and the exchange \mbox{$e$-$h$} repulsion. 
The term $\hat{H}_{eh}^{\rm II}= \delta \hat{H}_{eh}^{\rm II} + {\rm H.c.}$ with
\begin{equation}
\label{HeheII} 
 \delta \hat{H}_{eh}^{\rm II} =
       \sum_{\substack{m_1 , m_2 \\ m_1' , m_2'}}
	   \sum_{\substack{s_1 , \sigma_1 \\ s_2 , \sigma_2}}
         \mathcal{W}_{N_1 m_1  \, N_2' m_2'}^{N_1' m_1' \, N_2 m_2}
                c^{\dag}_{N_1' m_1'} c^{\dag}_{N_2' m_2'} d^{\dag}_{N_2 m_2} c^{\vphantom{\dag}}_{N_1 m_1}
\end{equation}
and
$N_1=\{1, s_1,\sigma_1 \}$, $N_1'=\{0, s_1,\sigma_1 \}$,
$N_2=\{-1, s_2,\sigma_2 \}$, and  $N_2'=\{0, s_2,\sigma_2 \}$
does not conserve the number of particles in each of the LL's involved.
It describes the resonant conversion between a three-particle trion state and a single electron state in the $n=1$ LL
and contributes $\sim E_0$.
These indicated interaction terms exhaust {\em all} Coulomb contributions $\sim E_0$
for the considered charged excitations at $\nu = -2 + \epsilon$.
For other excitations and at other filling factors, the effective interaction Hamiltonians are constructed analogously.

The excitations may additionally be characterized by their total spin projection $S_z$
and  pseudospin projection (valley index) $\Sigma_z$ \cite{CasGPNG09,AbeABZ10}.
However, there exists a larger \SU group, which includes the direct product
of the spin and pseudospin groups SU$_2$\,$\otimes$\,SU$_2$\,$\subset$\,\SU as a subgroup.
More precisely, the Hamiltonian of the Coulomb (and in fact any long-range) inter-particle interactions is \SU symmetric up to small
symmetry-breaking terms \cite{GoeMD06}.
The dynamical \SU symmetry in graphene is based therefore
on the equivalence of the two valleys ($\Uparrow$,$\Downarrow$) and the two spin states ($\uparrow$,$\downarrow$),
including the {\em continuous} interchange between spins and valleys.
The four basis states (flavors) that form a fundamental representation of the \SU group
are $\downarrow\Downarrow$,
    $\downarrow\Uparrow$,
	$\uparrow\Downarrow$, and
	$\uparrow\Uparrow$.
We shall label them in analogy with the four flavors of down, up, strange, and charm quarks, by
$ \{ \downarrow\Downarrow, \downarrow\Uparrow, \uparrow\Downarrow, \uparrow\Uparrow \} \equiv \{ d, u, s, c \}$.
The generators of the SU$_4$ group can be constructed using the bilinear combinations conserving
the number of electrons and holes. We have
$C_{ij} =\sum_{\mathcal{N}} c^{\dagger}_{\mathcal{N}i} c_{\mathcal{N}j}
       - \sum_{\mathcal{N}}d^{\dagger}_{\mathcal{N}j} d_{\mathcal{N}i}$.
 	   where the composite index
	   $\mathcal{N}=\left\lbrace n,m\right\rbrace $ and $i,j\in\left\lbrace d,u,s,c \right\rbrace$.
The generators satisfy the commutation relations
$[C_{ij}, C_{kl} ] = \delta_{jk} C_{il} - \delta_{il} C_{kj}$.
The operators of total spin $\hat{ \mathbf{S} }$ and pseudospin $\hat{ \bm{\Sigma} }$
can be expressed in terms of the $C_{ij}$.
For example,
$\hat{\Sigma}_z =  \frac{1}{2}\left( C_{cc} + C_{uu} - C_{ss} - C_{dd}\right)  $
and
$\hat{\Sigma}_+ = \hat{\Sigma}_x + i \hat{\Sigma}_y= C_{ud} + C_{cs}$.
The degeneracy of the eigenstates follows from the invariance of the Hamiltonian
 $\hat{H}=U^{\dagger} \hat{H} U $, where
$U=\exp \left( i \sum_{ij }\Theta_{ij} C_{ij} \right)$, with $\Theta_{ij}$ the \SU transformation ``angles'' \cite{GreM97,GoeMD06}.

To determine the multiplicities of the $64$ possible $eeh$ flavor states,
we need to decompose
%
%
the direct product of the three SU$_4$ multiplets.
The electron states form the fundamental \SU $[4]$-multiplet
with four states represented by the Young diagram  $\tiny{\yng(1)}$.
The hole (antiparticle) transforms as a conjugate
multiplet $[\bar{4}]$ and is represented by the Young diagram $\tiny{\yng(1,1,1)}$, which is one
box short of the unitary singlet  $\tiny{\yng(1,1,1,1)}$. The latter transforms as the \SU vacuum.
The decomposition of  the direct product of the SU$_4$ $eeh$ states gives
$[4] \otimes [4] \otimes [\bar{4}] = [4] \oplus [4] \oplus  \left[ \smash{\overline{20}} \right] \oplus [36]$.
The corresponding Young diagrams are, respectively,
\begin{equation}
\yng(1) \otimes \yng(1) \otimes \yng(1,1,1) = \yng(2,1,1,1) \oplus \yng(2,1,1,1)
\oplus \yng(2,2,1) \oplus \yng(3,1,1)
\; .
\label{eq-young}
\end{equation}
Graphically, the resulting multiplets with their flavor contents resolved are shown in
Fig.~\ref{fig-bound}(b)-(d). There, we use the standard elementary
particle physics \SU representation \cite{GreM97,multiplet}. Namely, the three orthogonal axes show:
the flavor isospin projection $\hat{T_z} = \frac{1}{2}\left( C_{uu} - C_{dd}\right) $,
the hypercharge $\hat{Y} = \frac{1}{3}\left(C_{uu}+C_{dd}+C_{cc} \right)-\frac{2}{3}C_{ss} $
and the charm $\hat{c} = C_{cc}$.
The Young diagrams \eqref{eq-young} show that the states in the
$\left[ \smash{\overline{20}} \right]$ multiplet
are antisymmetric with respect to permutation
of two electrons and symmetric in the $[36]$ multiplet.
Notice that the states of the two $[4]$-multiplets describe the combination
of an electron in one of the possible $d, u, s, c$ states
with an  $eh$ symmetric neutral state involving the other electron
$Q_0 \equiv \frac{1}{2} \left( d\bar{d} + u\bar{u} +  s\bar{s} + c\bar{c} \right)$;
the two electrons are in a mixed state, {\it i.e.}, do not have any specific permutation
symmetry with respect to one another. The two flavor $[4]$-multiplets
are combined with their conjugate coordinate parts to form an antisymmetric state
as dictated by Fermi statistics.
The state $Q_0$ is the flavorless unitary singlet $\tiny{\yng(1,1,1,1)}$, which carries no \SU
quantum numbers (all zeros).
This state is also the spin- and isospin-singlet with $S=0$ and $\Sigma=0$.

\begin{figure}[tbh]
  \centering
\includegraphics[width=0.85\columnwidth]{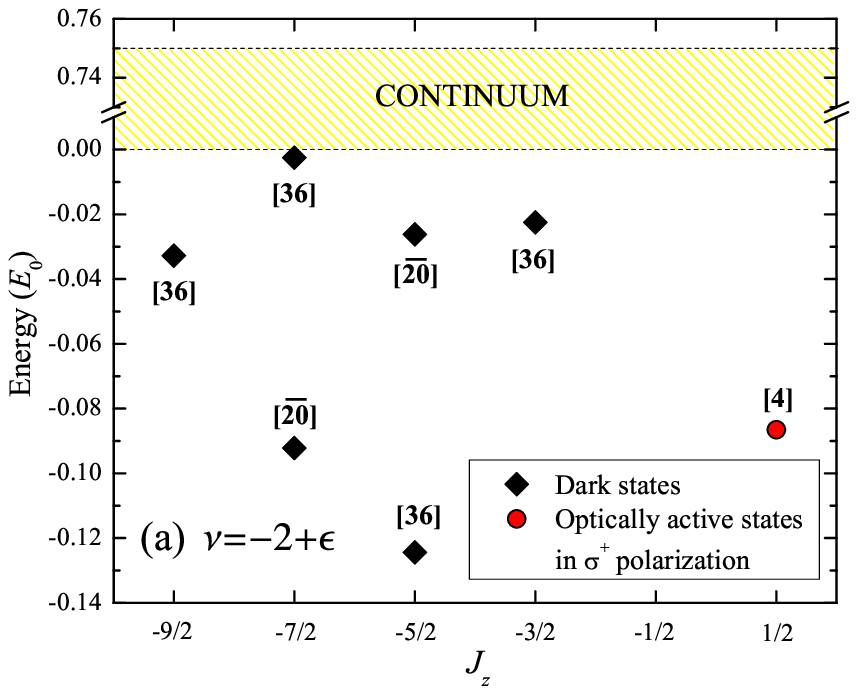}\\
\includegraphics[width=0.85\columnwidth]{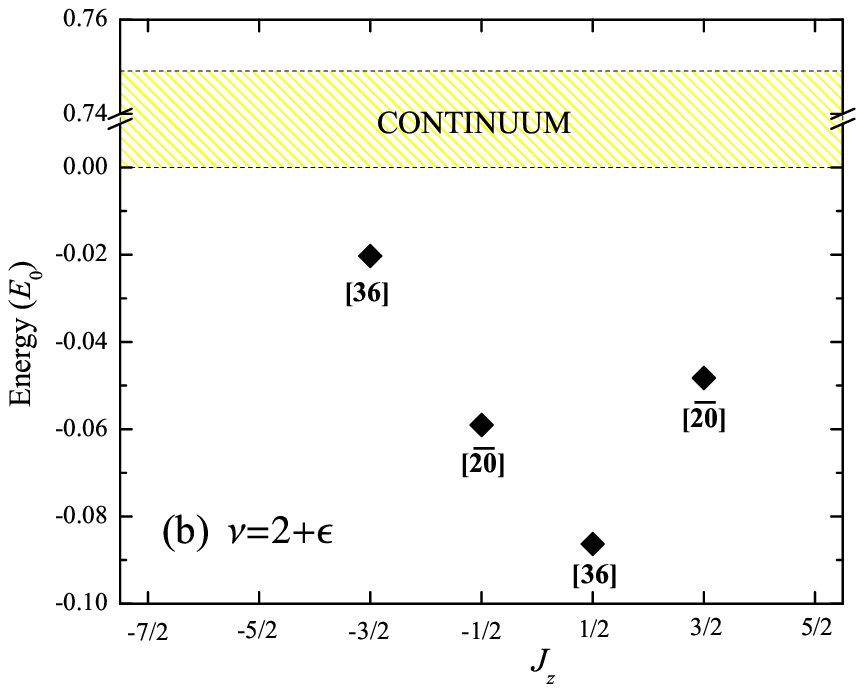}
  \caption{(color online) Bound $X^-$ states and their multiplicities for (a) $\nu= -2+\epsilon$ and (b) $\nu= 2+\epsilon$.
  Only the seed $k=0$ states are shown. Energy zero corresponds to the CR mode at $\hbar\omega_c$.
  The absorption spectrum of the bright state at $J_{z}=1/2$ with energy $-0.09 E_0$ is shown in Fig.~\ref{Optics}.
  }
\label{fig-enu}
\end{figure}

The structure and the multiplicities of the obtained bound trions $X^-$ are shown
for filling factors $\nu = \pm 2 + \epsilon$ in Fig.~\ref{fig-enu}.
We truncate the basis in Eq.~(\ref{eq-Rdagger}), so that only single particle oscillator quantum numbers $m\le 15$ are used
and diagonalize the Hamiltonian matrices
numerically.
The latter typically include about 800 basis states, depending on the particular value of $J_z$.
The form of the Coulomb matrix elements and other numerical details were presented in \cite{FisDR09a}.
We performed calculations for trions with no spin or pseudospin flips for the filling factors $\nu=\mu\pm\epsilon$ with $\mu=-2,-1,\hdots,2$.
Stable bound $X^-$ states were found only for $\mu=\pm2$.
Our results are valid in the presence
(as indicated in Fig.~\ref{fig-bound}) as well as the absence of
spin $\delta \hbar \omega_{s}$ and valley  $\delta \hbar \omega_{v}$ LL splittings,
provided $\delta \hbar \omega_{s,v} \ll E_\mathrm{0}$.
In very strong magnetic fields $B > 20$\,T, the $n=0$ LL splitting may become
 of the order of $0.1 E_0$ \cite{JiaZSK07}, most probably due to the spontaneous symmetry breaking \cite{CasGPNG09,AbeABZ10}.
If this occurs, the \SU multiplets will split, which requires additional group theoretical analysis.

The \SU symmetry allows one to determine, in a very direct and simple manner,
the selection rule for trion photocreation
$\hbar\omega + e^- \rightarrow (eeh)$.
Indeed, the interaction of electrons with light of frequency $\omega$ and
circular polarizations $\sigma^{\pm}$ is described by the Hamiltonian
$ \delta \hat{H}_{\pm} =  {\cal A}
                \left( \begin{smallmatrix}
            \tau_{\pm}     &   0   \\
            0     &    \tau_{\pm}
                       \end{smallmatrix} \right)$,
where $\tau_{\pm} = \tau_x \pm i \tau_y$
are the isospin Pauli matrices acting in the space of the graphene sublattices,
${\cal A} = \frac{e v_F \mathcal{E}e^{-i\omega t}}{i\omega c}$
and $\mathcal{E}$, the electric field amplitude \cite{FisDR09a,AbeF07,GusSC07}.
When expressed in the \SU terms, $ \delta \hat{H}_{\pm}$ assumes the form
${\cal A} \left( d\bar{d} + u\bar{u} +  s\bar{s} + c\bar{c} \right) \sim Q_0$,
{\it i.e.\/}, is proportional to the \SU singlet $Q_0$.
This means that the photon is flavorless in the \SU sense.
From this and considerations of the dipole transition matrix elements
$\langle (eeh)| Q_0 |e^-\rangle$,
it immediately follows that the states
in the $\left[ \smash{\overline{20}} \right]$  and $[36]$ multiplets are all dark
while the states in the $[4]$-multiplet may be bright,
provided they have proper orbital quantum numbers.
Let us discuss the latter.

The optical selection rules for the orbital quantum numbers
are $\Delta k =0$ and $\Delta J_z = \pm 1$ in the $\sigma^{\pm}$ polarizations.
The former reflects the negligible linear momentum of a photon in the dipole approximation
and the fact that  $k$  is associated with translations
({\it i.e.\/}, is the discrete analog of linear momentum for charged states in a magnetic field).
Let us consider $X^-$ photocreation
$\hbar\omega_{\sigma^{\pm}} + e^-_{n m} \rightarrow (eeh)_{k \, J_z\hspace{-1pt}-\hspace{-0.3pt}k}$;
here and in what follows $J_z$ is that of the seed state.
The electron in the initial state in the $n$-th LL
has $j_z = |n| - m - \frac{1}{2}$,
and $\Delta k =0$ reads in this case $k=m$.
Therefore, only the $X^-$  states from the  \SU $[4]$-multiplet
having $J_z = |n| + \frac{1}{2}$ for $\sigma^+$
and $J_z = |n| - \frac{3}{2}$ for $\sigma^-$ can be photocreated.
We see that the combination of the \SU and orbital selection rules imposes
stringent limitations on optical transitions.

As a result, among all of the bound $X^-$ states found near the charge
neutrality point,
the only bright states are at $\nu = -2 + \epsilon$ with the binding energy
$\sim 0.1 E_0$  ($J_z = \frac{1}{2}$, active in $\sigma^+$, see Fig.~\ref{fig-enu}(a)).
Depending on the broadening, this may give rise to a separate absorption peak below the CR
or may show up as a low-energy CR shoulder. This is shown in Fig.~\ref{Optics},
where the spectra  are broadened $\sim 0.05 E_0$, which amounts to 2\,meV at $B=10$\,T.
We conclude that observation of the bright $X^-$ feature is almost feasible in graphene
even for the current broadening/disorder experimental parameters such as mobilities of $\sim 17000$ cm$^2$/Vs \cite{JiaZSK07,AbaSDA10,HenCJL10}. We may compare the binding energy to that found for trions in the conventional two-dimensional electron gas (2DEG). This is also $\sim 0.1 E_0$ \cite{DzyS00}, but $E_0$ is smaller than it is in graphene, due to a larger value of the dielectric constant $\varepsilon$ for the 2DEG.

\begin{figure}[t]
  \centering
  \includegraphics[width=0.75\columnwidth]{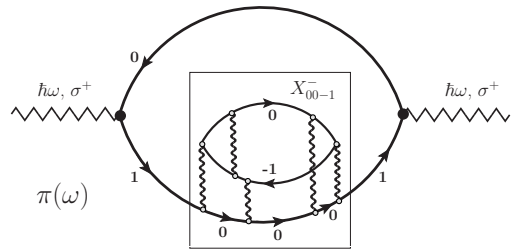}
  \caption{Photon polarization operator $\pi(\omega)$.
  The box shows the resonant excitation of the \mbox{$e$-$h$}
  pair with the deexcitation of the electron to $n=0$ LL.
  Because of the multiple Coulomb interactions (wiggly vertical lines) the intermediate state
  becomes a bound trion $X^-_{00-1}$, with two electrons in the $n=0$ LL and a hole in the $n=-1$ LL,
  leading to a singularity in $\pi(\omega)$.
  }
   \label{dressing}
\end{figure}

The oscillator strength of the bright $X^-$ with $J_z = \frac{1}{2}$ needs some clarification.
Indeed, the {\em hole-like\/} collective excitations at $\nu = -2 + \epsilon$
originate in the lower cone (inset in Fig.~\ref{Optics}).
As such, they are optically inactive in the electron-like $\sigma^+$ polarization.
This holds, however, only in the lowest order in the Coulomb interactions \cite{FisDR09a}.
In fact, the resonant $e_{1} \rightleftarrows X^-_{00-1}$ processes described by the Hamiltonian $\hat{H}_{eh}^{\rm II}$ and indicated in Fig.~\ref{dressing}
lead  to ``dressing'' of the optical response.
As a result, the oscillator strength of the bright $X^-$ becomes comparable to that
of the CR mode per electron.
Thus, the intensity of the $X^-$ peak relative to the main CR  is proportional
to the number of excess electrons $\epsilon$ \cite{FinSB95,BuhMWB95,AstYRC05,ShiOSP95}.
Notice that  because of the $e$-$h$ symmetry, there exists the $X^+$
counterpart at $-\nu =  2 - \epsilon$, which is bright in $\sigma^-$  polarization.

The dark bound $X^-$ and $X^+$ states may in principle be observed by tunneling spectroscopy,
which is a sensitive tool for probing discrete states in the spectrum. This can be pursued, e.g.,
in tunneling experiments involving gate-tunable graphene quantum dots \cite{GueFSI10}.
Also, the various symmetry-breaking terms such as external fields, disorder and
lattice defects, ripples, and deformations \cite{CasGPNG09,AbeABZ10}
may partially lift the limitations following from the
orbital and \SU selection rules. This can make some of the dark states ``gray'' and
detectable in the absorption spectra as well as possibly in photoluminescence \cite{SchBSH03}.

In conclusion, we considered Coulomb correlations in graphene with a low density of excess electrons or holes
in partially filled LL's. We have shown that bound negative $X^-$ and positive $X^+$
trions exist in the spectra. With the \SU symmetry of graphene these are
analogous to quark-quark-antiquark complexes.
They are different from their counterparts in the conventional 2DEG systems such as GaAs \cite{FinSB95,BuhMWB95,AstYRC05,ShiOSP95}.
We expect that the predicted additional absorption peak due to the bright $X^-$
below the CR mode should undergo an evolution
with filling factor $\epsilon$ of a partially filled level.
This collective excitation can serve as an optical probe as its behavior should reflect correlations in the electron
system \cite{SchBSH03,NicYDM02}.
This evolution would be especially interesting to follow and study in the vicinity of
$\epsilon =\frac{1}{3}$ and other FQHE fractions.

After completion of this work we became aware of
publication \cite{BosSSH10}, where the observation of a plasmaron excitation, a zero field analogue of the $X^+$,
has been reported.

\begin{acknowledgments}
We thank Paul Harrison for discussions.
We are grateful to EPSRC (Warwick) and the Cottrell Research Corporation (CSUB) for funding.
ABD gratefully acknowledges the Scholarship of KITP, UC Santa Barbara.
\end{acknowledgments}


\begin{thebibliography}{10}
\bibitem{quarks}
The dynamical interactions, strong for quarks and electromagnetic for electrons, are of course fundamentally different.
Besides, electrons in graphene lack the color of true quarks.
The latter makes the existence of states analogous to $qq\bar{q}$ and $q\bar{q}\bar{q}$ possible in graphene,
while such states cannot be colorless and, as such, do not exist in elementary particle physics.
\bibitem{multiplet}
The $qq\bar{q}$  multiplet $\left[ \smash{\overline{20}} \right]$
is conjugate to the baryon $qqq$ $[20]$-plet of elementary particle physics
including the neutron $n$, proton $p$  and $\Lambda$, $\Sigma$, and $\Xi$ particles.
\end{thebibliography}

\end{document}